\documentclass[american,aps,twocolumn,showpacs]{revtex4-2}
\usepackage[latin9]{inputenc}
\usepackage{dcolumn}
\usepackage{color}
\usepackage{textcomp}
\usepackage{amsmath}
\usepackage{amssymb}
\usepackage{graphicx}
\usepackage{braket}

\makeatletter

\usepackage{xcolor}
\usepackage{dcolumn}
\usepackage{bm}
\usepackage{wrapfig}
\usepackage{subfigure}
\usepackage{ucs}
\usepackage{lineno}
\usepackage{epsfig}
\usepackage{psfrag}\usepackage{epstopdf}
\DeclareGraphicsExtensions{.pdf,.eps,.png,.jpg,.mps}

\usepackage{babel}

\makeatother

\usepackage{babel}
\begin{document}

\title{Study of the ground state and thermodynamic properties of Cu5-NIPA-like molecular
       nanomagnets}

\author{J. Torrico$^1$ and J. A. Plascak$^{1,2,3}$ }

\affiliation{$^1$  Departamento de Física, Universidade Federal de Minas
Gerais, C. P. 702, 30123-970, Belo Horizonte-MG, Brasil}
\affiliation{$^2$ Departamento de F\'{\i}sica, Universidade Federal da Para\'{\i}ba,
Caixa Postal 5008, 58051-900, Jo\~ao Pessoa-PB, Brazil}
\affiliation{$^3$University of Georgia, Department of Physics and Astronomy,
30602 Athens-GA, USA}

\begin{abstract}
The thermodynamic properties of a spatially anisotropic spin-1/2 Heisenberg model for a
Cu$_5$ pentameric molecule is studied through exact diagonalization. The elementary
geometry of the finite lattice is defined on nanomolecules consisting of an hourglass
structure of two corner-sharing scalene triangles which are related by inversion symmetry.
This microscopic magnetic model is quite suitable to describe the molecular nanomagnetic
compound Cu5-NIPA. The ground-state phase diagram, as well as the
corresponding total magnetization, are obtained as a function of the anisotropic
exchange interactions and the external magnetic field. The thermodynamic behavior
of the model at finite temperatures is also studied and the corresponding
magnetocaloric effects are analyzed for various values of the Hamiltonian parameters.
\end{abstract}

\maketitle

\section{Introduction}
\label{intro}

The experimental and theoretical studies of the intrinsic magnetism in large molecules
and molecular nanomagnets have immensely increased in the last few decades \cite{coro,gat}.
One of the main reasons for the great deal of appeal in these molecular nanomagnets
resides in the possibility to experimentally access a wide variety  of quite interesting
fundamental quantum mechanical effects, such as quantum tunneling of the magnetization
\cite{gunt,tom,chio}, quantum phase interference \cite{wern}, level crossings and
magnetization plateaux \cite{taft,juli}, among others. What is even more appealing is the fact
that these compounds turn out to be promising materials for several practical applications
including, for instance, spintronics and quantum computing \cite{coro}. In addition, some
of these molecular magnets exhibit also magnetocaloric effects \cite{tishin,rome}, which
could make them highly suitable in designing a new era of magnetic refrigerators \cite{rome},
operating not only at everyday room temperatures and temperatures of hydrogen and helium
liquefaction (20 - 4.2 K), but also at subkelvin intervals needed in
important low temperature experimental devices.

One of these highly interesting experimental realization of modern nanomagnetic materials
is the chemical compound Cu$_5$(OH)$_2$(NIPA)$_4\cdot$10H$_2$O, short named as Cu5-NIPA,
with 5-NIPA standing for 5-nitro-isophtalic acid ligand. This system presents a rather
uncommon pentameric form of a triangle-based structure, with a pair of corner-sharing scalene
triangles, possessing an hourglass-like geometry. Some of the static magnetic properties
of Cu5-NIPA molecules, including their spin dynamics behavior, have been investigated
by Nath et al. \cite{nath}. Based on experimental results and ab initio density-functional
band-structure calculations, a microscopic theoretical Heisenberg model,
with spatially anisotropic exchange interactions, has been proposed
to describe Cu5-NIPA. It has been shown that the Cu ions carry a spin-1/2 and, despite the
triangular structure and ferromagnetic and antiferromagnetic interactions, the couplings
lead to no magnetic frustation. This is in contrast, for example, to V6 like
nanomolecules, where two independent triangles interact with exchange interactions
causing molecular spin frustation \cite{muller,luban,haraldsen,kowa,jordana}.

More recently, Sza\l owski and Kowalewska \cite{karol} have studied the ground-state
spectrum and the thermodynamic properties, including the magnetocaloric effects, of
these molecular nanomagnets. They have considered, however, only the experimental exchange
interaction values derived in Ref. \cite{nath} for the triangle bonds present in the Cu5-NIPA
molecule. It would be thus quite interesting to see what will be the molecule behavior by
considering different values for the coupling interactions, including sign change of the
original ferromagnetic one, which will induce some magnetic
frustration. As in ordinary magnetic materials, such variation of the exchange
interactions could be physically plausible by controlling, for instance, some external
conditions like the hydrostatic pressure. By using the exact diagonalization
of the Hamiltonian, it will be shown that this molecule has indeed a very rich ground state phase
diagram as a function of the model parameters, specially for the more symmetric case.
Accordingly, the thermodynamics at finite temperatures also conveys interesting magnetocaloric
effects, in its direct and inverse forms. We will here use a similar procedure recently employed in
the study of V6 nanomolecules of Ref. \cite{jordana}.

The outline of this paper is as follows. In the next section, we describe the Hamiltonian
model for the Cu5-NIPA-like molecules. In section \ref{eigenen} we present the exact
analytical diagonalization for the energy spectrum and the corresponding more relevant
eigenstates. The general behavior of the energy spectrum and the ground state phase diagram
are discussed in section \ref{phasdia}. In section \ref{thermo} we investigate
the magnetic and thermodynamic properties of the system through the study of the
magnetization, susceptibility, entropy, magnetocaloric effect and specific heat.
Finally, some concluding remarks are drawn in the last section.

\section{Hamiltonian model}
\label{model}

\begin{figure}[h]
\includegraphics[scale=0.32]{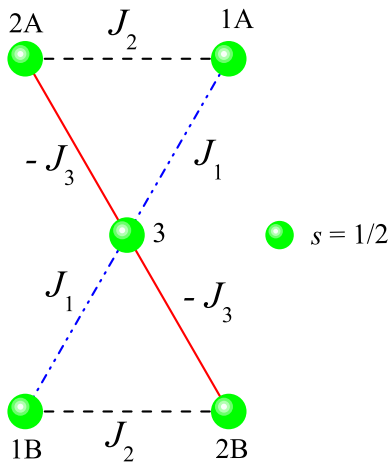}
\caption{\label{figmol} (color online) Schematic illustration of a Cu5-NIPA molecular
 magnet. The full circles represent the Cu$^{2+}$ ions with spin-1/2. $A$ and $B$ denote
 the two triangles, with sites 1$A$, 2$A$, and 1$B$, 2$B$, respectively, both sharing the
 common site labeled as 3. $J_1$, $J_2$ and $J_3$ are the corresponding exchange interactions.}
\end{figure}

The Cu5-NIPA-like magnetic molecules have five Cu$^{2+}$ ions, carrying each one a spin-1/2,
which are located at the vertices of two corner-sharing triangles, forming an hourglass structure,
with inversion symmetry, as is schematically depicted in Fig. \ref{figmol}.

Based on the experimental results of Ref. \cite{nath}, the corresponding Hamiltonian
for this molecule can be written as
\begin{eqnarray}
  \mathcal{\hat{H}}&=&J_1( \vec{S}_{1A} \cdot \vec{S}_{3}+\vec{S}_{1B} \cdot \vec{S}_{3}) +J_2(\vec{S}_{1A} \cdot \vec{S}_{2A}+\vec{S}_{1B} \cdot \vec{S}_{2B})\nonumber \\
  ~&-&J_3(\vec{S}_{2A} \cdot \vec{S}_{3}+\vec{S}_{2B} \cdot \vec{S}_{3})\nonumber\\
  ~&-&B(S_{1A}^z+S_{2A}^z+S_{1B}^z+S_{2B}^z+S_{3}^z),
  \label{ham}
\end{eqnarray}
where $J_1$, $J_2$ and $J_3$ are the exchange interactions and $B$ is the external
magnetic field applied in the $z$ direction. $\vec{S}=(S_i^x, S_i^y, S_i^z)$ is the Heisenberg spin-1/2 operator with the
components $S_i^\alpha$ ($\alpha=x,y,z$) given by the Pauli spin matrices.

Specifically for the Cu5-NIPA molecule all the exchange interaction parameters in Hamiltonian
(\ref{ham}) are positive and are given by $J_1 = 217$ K and $J_2=J_3 = 62$ K, which results
in a non-frustrated molecule, despite the triangle structure with some antiferromagnetic
interactions.  In this work, we will study the effects of different exchange interactions in
this kind of molecule, allowing still for the presence of frustration with negative values
of $J_2$ and $J_3$, and analyzing the corresponding effects in the zero temperature phase
diagram as  well as in the thermodynamic behavior of the system.

\section{Eigenenergies and eigenstates}
\label{eigenen}

Using the eigenstates of the $z$-component $S^z_i$ spin operator that spam the Hilbert space of
$\cal H$, the above Hamiltonian is given by a $32 \times 32$ matrix that can be exactly
diagonalized by resorting to some modern technical computing system like Maplesoft.
This procedure has been previously done in Refs. \cite{nath,karol}. However, as in the
present study we will consider general values for the exchange interactions and external
magnetic field, the 32 eigenvalues so obtained are explicitly given in the Appendix.

Concerning the eigenstates, even if we omit those coming from the cumbersome solution of the
cubic equation (\ref{ce}), the remainder ones are still rather lengthy to be reproduced in
the Appendix. For this reason, we will present below only the corresponding eigenstates of
the more relevant low-lying eigenenergies, and grouping them according to their total
spin values $S$, since it is the quantity $S$ that gives the main magnetic behavior of the
molecule.

\subsubsection{Ferromagnetic state with $S=5/2$}

There is one state with all spins aligned with the magnetic field resulting in a molecular
total spin $S=5/2$. This ferromagnetic (FM) state has energy $E_{FM}$ and eigenstate
$\ket{FM}$, which are respectively given by
\begin{align}
  E_{FM}=\varepsilon_{1}=\frac{1}{2}J_1+\frac{1}{2}J_2-\frac{1}{2}J_3-\frac{5}{2}B,
\end{align}
\begin{align}
  |FM\rangle=|\uparrow,\uparrow,\uparrow,\uparrow,\uparrow\rangle ,
\end{align}
where an up arrow represents the $z$-component of the spin aligned in the same direction
of the external magnetic field (a down arrow means the opposite). This will be
the stable phase for large magnetic fields.

\subsubsection{Ferrimagnetic state with $S=3/2$}

This ferrimagnetic (FI1) state has spin-3/2 with eigenenergy $E_{FI1}$ and eigenstate
$\ket{FI1}$ given by
\begin{equation}
  E_{FI1}=\varepsilon_{11}=-\frac{1}{4}J_1+\frac{1}{4}J_3-\frac{3}{2}B
  -\frac{1}{4}r,
\end{equation}
\begin{align}
  |FI1\rangle=&\dfrac{1}{\sqrt{2a^2+b^2+2}}[|\uparrow,\uparrow,\uparrow,\downarrow,\uparrow\rangle\nonumber\\
  &+|\downarrow,\uparrow,\uparrow,\uparrow,\uparrow\rangle-a|\uparrow,\uparrow,\uparrow,\uparrow,\downarrow\rangle\nonumber\\
  &-a|\uparrow,\downarrow,\uparrow,\uparrow,\uparrow\rangle+b|\uparrow,\uparrow,\downarrow,\uparrow,\uparrow\rangle],
\end{align}
where
\begin{align}
  a=&\dfrac{3J_1+3J_3+r}{2(J_2+2J_3)}\\
  b=&\dfrac{3J_1-2J_2-J_3+r}{J_2+2J_3}\\
  r=&\sqrt{(J_1+J_3)^2+4J_2^2+8[J_1^2+J_3^2-J_2(J_1-J_3)]}.
\end{align}

\subsubsection{Ferrimagnetic state with $S=1/2$}

There is another ferrimagnetic state (FI2) with smaller total spin value given by
\begin{align}
  E_{FI2}=&\varepsilon_{31}=x_3-\frac{1}{2}B,
\end{align}
\begin{align}
  |FI2\rangle=&~c_1|\uparrow,\downarrow,\uparrow,\uparrow,\downarrow\rangle
  +c_2|\uparrow,\downarrow,\uparrow,\downarrow,\uparrow\rangle \nonumber \\
  +&~c_3|\downarrow,\uparrow,\uparrow,\downarrow,\uparrow\rangle
  +c_4|\downarrow,\uparrow,\downarrow,\uparrow,\uparrow\rangle,
\end{align}
where $x_3$ is the smallest solution of the cubic equation (\ref{ce}) and $c_i$, $i=1,2,3,4$,
are constants satisfying  the normalization condition $\sum_{i=1}^4c_i^2=1$. Due to the
awkward solutions of the cubic equation, this state, as well as the degenerated one described
below, have been numerically computed for general values of the exchange interactions.

\subsubsection{Degenerate ferrimagnetic state with $S=1/2$}

This phase (FID) occurs in the ground state only for $J_3=-J_1$, where one has
\begin{align}
  E_{FID}=\varepsilon_{21}=\varepsilon_{27}= -J_1-\frac{1}{2}J_2-\frac{1}{2}B,
\end{align}
with two different eigenstates
\begin{eqnarray}
  |FID_1\rangle=\dfrac{1}{\sqrt{6}}[ &-&|\uparrow,\uparrow,\downarrow,\uparrow,\downarrow\rangle
  +|\uparrow,\uparrow,\downarrow,\downarrow,\uparrow\rangle \nonumber \\
  &-&\left.|\uparrow,\downarrow,\uparrow,\uparrow,\downarrow\rangle
  +|\uparrow,\downarrow,\downarrow,\uparrow,\uparrow\rangle\right. \nonumber \\
  &-& |\downarrow,\uparrow,\uparrow,\uparrow,\downarrow\rangle
  +|\downarrow,\uparrow,\uparrow,\downarrow,\uparrow\rangle],
\end{eqnarray}
and
\begin{eqnarray}
  |FID_2\rangle=\dfrac{1}{\sqrt{6}}[&+&|\uparrow,\uparrow,\downarrow,\uparrow,\downarrow\rangle
  -|\uparrow,\uparrow,\downarrow,\downarrow,\uparrow\rangle \nonumber \\
  &-&|\uparrow,\downarrow,\uparrow,\downarrow,\uparrow\rangle
  + |\uparrow,\downarrow,\downarrow,\uparrow,\uparrow\rangle \nonumber \\
  &-&|\downarrow,\uparrow,\uparrow,\uparrow,\downarrow\rangle
  +|\downarrow,\uparrow,\downarrow,\uparrow,\uparrow\rangle],
\end{eqnarray}
where, in this case, the smallest eigenvalue of the cubic equation has been labeled as $x_1$.
Note that for $J_3=-J_1$ the molecule is more symmetric resulting in a much simpler analytical
expression for the eigenvalues and eigenstates.

\section{Energy spectrum and ground state phase diagram }
\label{phasdia}

In what follows, we will fix our energy scale by assigning the value $J_1=1$ or, in other
words, measuring the other exchange interactions, including the magnetic field, in units of
$J_1$.

\subsection{Energy spectrum}

Fig. \ref{spec} shows the complete set of energy levels, which is formally given in the Appendix,
as a function of the external magnetic field $B$, for different values of the exchange
interactions $J_2$ and $J_3$. The most relevant low-lying states for our purpose are highlighted
according to the legend on top of panels (a) and (b). One can see that, for large values of the
external magnetic field, the stable phase is the ferromagnetic one, regardless the strength of the
exchange interactions. For $J_2=J_3=1$ in Fig. \ref{spec}(a), as the external field increases
from zero, the molecule undergo a first-order phase transition from the state FI2 to FI1,
following another phase transition from FI2 to FM. Note that as $B$ increases, so does the total
molecular spin $S$, the latter one jumping from 1/2 to 3/2 in the first transition, and from 3/2
to 5/2 in the second transition, as should be expected.

\begin{figure}[h]
\includegraphics[scale=0.4]{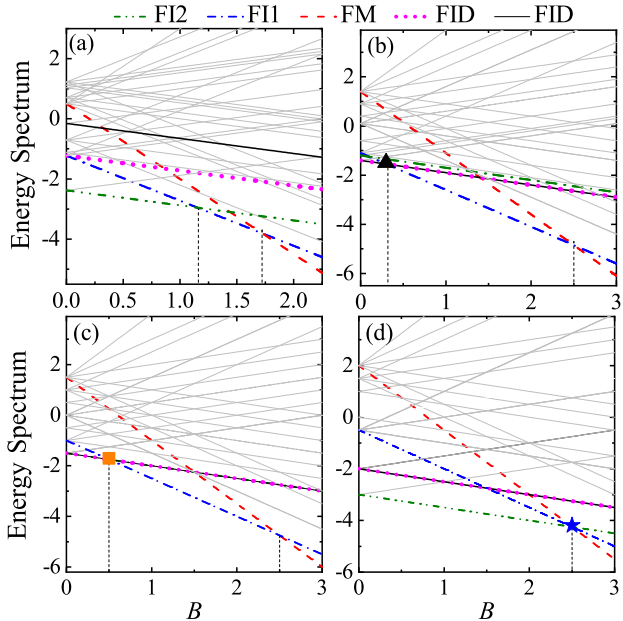}
\caption{\label{spec} (color online)
 Energy spectrum as a function of the external magnetic field $B$, for some values of the
 exchange interactions $J_2$ and $J_3$. We have in (a) $J_2=1$ and $J_3=1$; in (b) $J_2=0.8$
 and $J_3=-1$; in (c) $J_2=1$ and $J_3=-1$ and in (d) $J_2=2$ and $J_3=-1$. The legend of the
 low-lying states are given on top of panels (a) and (b). The energy of both FID phases
 are the same in (b)-(c), and are also equal to the energy of FI2 phase for $J_2=1$ in (c).
 The triangle in (b), the square in (c) and the star in (d) represent triple, quadruple and
 quintuple points, respectively. All other crossings are ordinary two-phase coexistence. }
\end{figure}

It is interesting to see what happens when $J_3=-1$, making the molecule more symmetric but
inducing, at the same time, a frustration in the ordering of the $z$-component spin of Cu
ions. This is shown in Figs. \ref{spec}(b), (c) and (d) for increasing values of $J_2$,
namely $J_2=0.8,~1$, and 2. For $J_3=-1$, the two FID phases have the same energies and are
the ground state for small fields for $J_2=0.8$. The triangle in Fig. \ref{spec}(b) signals
the triple point where the molecule transitions to the FI2 phase. However, for $J_2=1$ the
FID phases also coexist with the FI2 phase in a triple line for low magnetic fields, before
transitioning to the FI1 phase at the quadruple point represented by the square in
\ref{spec}(c). For $J_2=2$, the FID and FI1 phases are suppressed as the lowest energies,
and only a transition from FI2 to the FM phase takes place. However, at this transition,
the energy of FI1 phase, as well as the eigenenergies $\epsilon_7$ and $\epsilon_9$
(which are not highlighted in the other panels because they are only significant for this
case), are all the same, turning this transition into a quintuple point, which is given
by the star in  \ref{spec}(d).

\subsection{Ground state phase diagram}

From Fig. \ref{spec} it is easy to see that we can construct the phase diagram, in terms of
the Hamiltonian parameters, by just seeking the crossing values of the energies of the states
FM, FI1, FI2 and FID. This is easily achieved by equating the corresponding low-lying
eigenenergies. As a matter of example, the transition from FI1 to the FM phase is given by
\begin{equation}
B=\dfrac{3}{4}J_1+\dfrac{1}{2}J_2+\dfrac{3}{4}J_3+\dfrac{1}{4}r.
\end{equation}
The other transition lines involve the numerical solutions of Eq. (\ref{ce}).

\begin{figure}[h]
\includegraphics[scale=0.45]{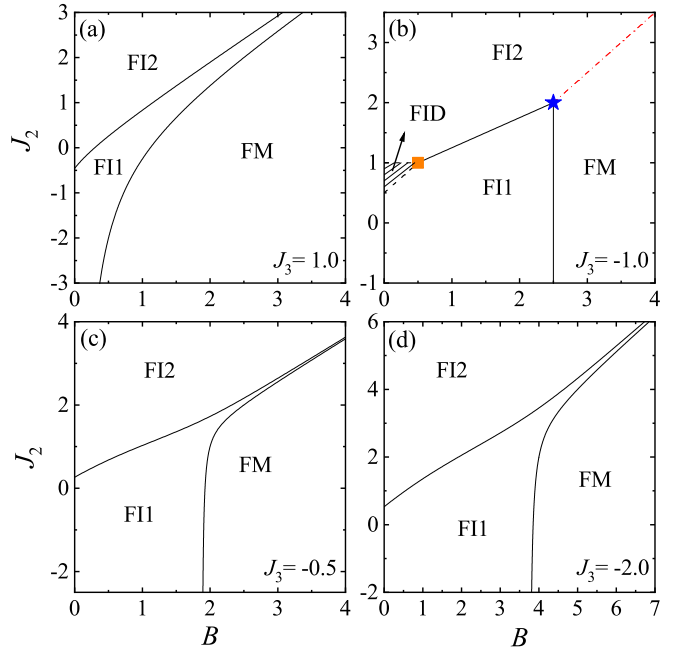}
\caption{\label{pd} (color online)
 Ground-state phase diagram in the $J_2$ versus $B$ plane. The full lines represent transitions
 where two phases coexist, the dashed lines where three phases coexist (triple line), and the
 dashed-dotted line has four phases coexisting (quadruple line). Accordingly, the square is a
 quadruple point and the star is a quintuple point. The hatched area is a region where the two
 FID phases coexist. The values of $J_3$ are given in the bottom right of each panel.}
\end{figure}

Fig. \ref{pd} depicts the topology of the phase diagram in the $J_2$ versus $B$ plane for
different values of the interaction $J_3$. For positive values of $J_3$, the phase diagrams
are quite similar to that shown in Fig.
\ref{pd}(a) for $J_3=1$. The FI1 phase lies always between the FI2 and FM phases and their
boundaries correspond to two-phase coexistence lines. High positive values of $J_2$
force the $z$-component of the spins in sites $1A$ and $2A$ to be antiferromagnetically ordered,
with the same ordering for the sites $1B$ and $2B$. For positive values of $B$ the total spin of
the molecule will then be $S=1/2$, the FI2 phase. On the other hand, for negative values of $J_2$,
a ferromagnetic ordering will prevail between sites $1A$ and $2A$, as well as between sites
$1B$ and $2B$, implying that for small values of $B$ the FI1 phase with $S=3/2$ will be stable
instead. As expected, for strong external magnetic fields the tendency is the molecule to be
in the FM phase.

The interesting case $J_3=-1$ is depicted in Fig. \ref{pd}(b), where one has a region of the
two FID phases coexisting for low magnetic fields. These two-coexisting phases are separated
from the single FI2 and FI1 phases by triple lines. These triple lines meet the two-phase
transition line from the FI2 to FI1 phases in a quadruple point. The transition line from
the FM phase to the FI2 phase is in fact a quadruple line, because in addition to the coexistence
of the FM and FI2 phase, we still have the eigenenergies $\epsilon_7$ and $\epsilon_9$ being
equal to the energies FI2 and FM in this parameter region. As a result, the common point of
the FI1, FI2 and FM phases in Fig. \ref{pd}(b) turns out to be a quintuple point. A more
detailed view of the coexisting phases at the above quadruple and quintuple points can be
seen in Figs. \ref{spec}(c) and (d), respectively.

For $J_3\ne -1$ the phase diagrams are similar to that shown in Fig. \ref{pd}(a), as can be
seen in Figs. \ref{pd}(c) and (d) for $J_3=-0.5$ and $J_3=-2$, respectively. The main difference
is the narrowing of the FI1 phase region, between the FI2 and FM phases, for large values of
the external magnetic field.

\section{Thermodynamic properties}
\label{thermo}

The thermodynamic properties of these magnetic molecules can be obtained by calculating
the partition function through
\begin{eqnarray}
  \mathcal{Z}=\sum_{i=1}^{32}\mathrm{e}^{-\beta \varepsilon_i},
  \label{z}
\end{eqnarray}
where $\beta=1/k_BT$, $k_B$ is the Boltzmann constant, $T$ is the absolute temperature,
and $\varepsilon_i$ are all the eigenvalues of the present Hamiltonian (which
are given in the Appendix). Although in the literature Eq. (\ref{z}) is referred to as the
canonical ensemble partition function, the proper designation for this ensemble
would be {\it field ensemble}, as discussed in Ref. \cite{pla}.
However, independent of the given ensemble name, the corresponding free energy per molecule
will be given by
\begin{eqnarray}
  F=F(T,B)=-k_BT\ln(\mathcal{Z}).
\end{eqnarray}

From the above free energy, we can compute all thermodynamic quantities of interest. In the
present case we will analyse the magnetization $M$ and
susceptibility $\chi$, entropy $\mathcal{S}$ and specific heat $C$, which can be
calculated from the well known thermodynamic relations
\begin{eqnarray}
  M&=&-\left(\dfrac{\partial F}{\partial B}\right)_T, \quad
  \chi=\left(\dfrac{\partial M}{\partial B}\right)_{T},\quad \\
  \mathcal{S}&=&-\left(\dfrac{\partial F}{\partial T}\right)_B, \quad
  C=T\left(\dfrac{\partial S}{\partial T}\right)_B.
\end{eqnarray}

An interesting quantity that can be studied in this system is the so called magnetocaloric
effect, which is defined by the adiabatic temperature change, or the isothermal entropy change, as
the  external magnetic field is varied. This effect can be quantified by the following relation
\begin{eqnarray}
  \left(\frac{\partial T}{\partial B}\right)_S=-\frac{(\partial S/ \partial B)_T}
  {(\partial S/ \partial T)_B}.
  \label{mceq}
\end{eqnarray}

Another quantity related to the magnetocaloric effect is the change of the magnetic entropy due to a
change in the magnetic field and can be written as
\begin{eqnarray}
  \Delta S(T,\Delta B)&=&\int_{B_i}^{B_f} \left(\dfrac{\partial M(T,B)}{\partial T}\right)_BdB
  \nonumber \\
  &=&S(T, B_f) - S(T,B_i),
  \label{mce}
\end{eqnarray}
where $\Delta B=B_f-B_i$, $B_i$ and $B_f$ being the initial and final magnetic fields,
respectively \cite{franco,pech}. When $-\Delta S>0$ the material has a conventional, or direct,
magenetocaloric effect (the system heats up), and when $-\Delta S<0$ the material presents
an inverse  magnetocaloric effect (the system cools down).

In what follows we will present the thermodynamic behavior of the model described by the
Hamiltonian (\ref{ham}). In all the results discussed below we considered $k_B=1$ in order to have
$k_BT/J_1\rightarrow T$. We have also taken the more interesting case $J_3=-1$, however, the
general trend of the results are qualitatively similar for other values of $J_3$.

\subsection{Magnetization}
\label{mag}

Looking at the ground state phase diagrams depicted in Fig. \ref{pd}, one can see that,
in general, for small values of $J_2$, mainly negative ones, one has only one first-order
transition from the $S=3/2$ to the $S=5/2$ state, while for higher values of $J_2$ one has two
transitions, since $S=1/2$ is also a stable phase for small values of $B$. Exception should
be made for $J_3=-1$ and $J_2>2$, where only one transition is seen from the $S=1/2$ to the
$S=5/2$ state. However, as discussed in the previous section, at this latter transition
there is also a coexistence with phases given by the eigenenergies $\epsilon_7$ and
$\epsilon_9$, where both correspond to eigenstates with $S=3/2$. Overall, at zero temperature,
the magnetization, as a function of $B$, has a stepwise shape with either two or three steps.

\begin{figure}[h]
\includegraphics[scale=0.43]{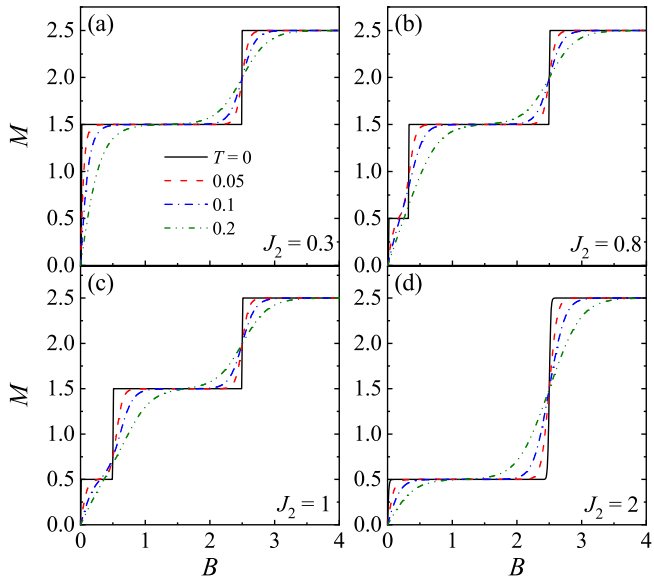}
 \caption{\label{mxb} (color online)
 Total molecular magnetization $M$, as a function of the magnetic field $B$, for $J_3=-1$
 and different values of temperature $T$ and exchange interaction $J_2$. The legend of the
 temperatures shown in (a) also applies to the other panels.}
\end{figure}

In Fig. \ref{mxb} we have the magnetization, as a function of external field, for $J_3=-1$ and
different values of the exchange interaction $J_2$ and temperature T.
In the ground state we have $M=S$ and the
magnetization has a stair style clearly showing the jumps of $M$ at the transitions. However,
as soon as we have a finite temperature, a continuous behavior takes place. At the same time,
the magnetization drops to zero at $B=0$ for any value of $T$. This is indeed an expected
behavior, because being the molecule a zeroth dimensional quantum system, it is equivalent to a
one-dimensional classical spin model, which in turn has no phase transition at finite
temperatures. However, it is worthwhile to see that for low temperatures, the system still
keeps a kind of memory of the ground state magnetization. The experimental results of the
measured magnetization of the Cu5-NIPA compound in Ref. \cite{nath} show exactly the above
behavior as the magnetic field increases from zero. Although only the transition from the
FI2 to the FI1 has been observed, the next transition to the FM phase should probably occur
for still higher values of the magnetic field.

\begin{figure}[h]
\includegraphics[scale=0.43]{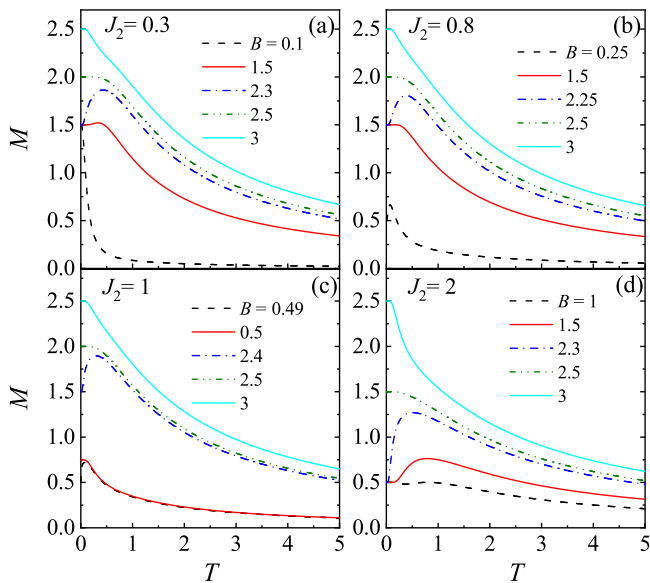}
\caption{\label{mxt} (color online)
 Total molecular magnetization $M$, as a function of temperature $T$, for $J_3=-1$ and different
 values of the external fields and exchange interaction $J_2$.
 }
\end{figure}

It is interesting to notice that the behavior of the magnetization, now as a function of the
temperature, has some peculiarities for external fields just smaller than the fields where
the jump occurs. For these values of $B$, the magnetization, as a function of temperature,
initially increases before smoothly start to decrease to zero. This is shown in Fig. \ref{mxt}
for several values of the external field (this figure takes the same exchange parameters as
Fig. \ref{mxb}). This unusual increase of $M$ with $T$ reflects the fact that, in this region,
the thermal fluctuations initially populate states having higher values of $S$.

For $J_3\ne-1$ the behavior of the magnetization is quite similar to those shown in Figs.
\ref{mxb} and \ref{mxt}.

\subsection{Susceptibility}
\label{suscep}

\begin{figure}[h]
\includegraphics[scale=0.43]{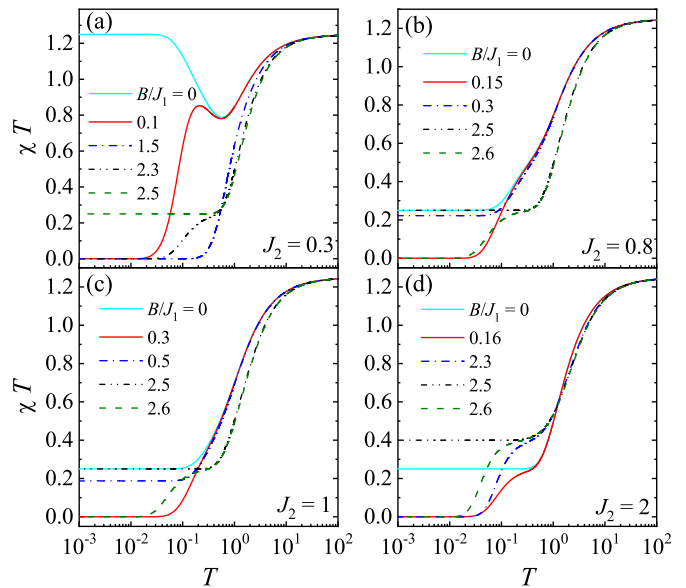}\\
 \caption{\label{sus} (color online) Susceptibility times
 temperature $\chi T$, as a function of temperature $T$ (in logarithmic scale), for $J_3=-1$
 and different values of the magnetic field $B$ and exchange interaction $J_2$.
 }
\end{figure}

The susceptibility times temperature, as a function of the temperature (in logarithmic scale
for a better visualization of the whole range of $T$),  is shown in Fig. \ref{sus} for $J_3=-1$
and various values of external field and exchange interaction $J_2$. For high values of
temperature one has the usual paramagnetic behavior for any values of the external field, as
expected. For low temperatures, one has $\chi T \rightarrow 0$ as $T\rightarrow 0$ for all values of
$B$, except those where the molecule magnetization changes its total
spin value at zero temperature. At these transition fields we can clearly see that
$\chi T \rightarrow constant$. Note, however, that this is not a quantum critical
phase transition, being only
an ordinary first-order transition (recall that for the one-dimensional Ising model one has
$\chi T \rightarrow \infty$ as $T\rightarrow 0$ \cite{baxter}).

As $T\rightarrow 0$ the susceptibility can be analytically obtained through an expansion of the
partition function at low temperatures. It turns out that, in this limit, $\chi$ is independent
of the exchange interactions $J_2$ and $J_3$, depending only on the constant prefactor multiplying
the magnetic field. Since $\chi T \rightarrow 1.25$ as $T\rightarrow 0$ for $B=0$ and $J_2=0.3$ in
Fig. \ref{sus}(a), and this value of the susceptibility is comparable to the value at high
temperatures, one can see a valley like behavior for intermediate values of $T$. On the other
hand, for $B=0.1$ the susceptibility should go to zero as $T\rightarrow 0$, in some sense
explaining the peak and valley present in Fig. \ref{sus}(a). In the other panels of Fig. \ref{sus},
the plateau at $T=0$ is responsible for the presence of the shoulder for external fields close
to the quantum phase transition ones.

It should be said that the experimental results of $\chi T$ obtained from Cu5-NIPA samples
\cite{nath} display indeed the general form shown in Fig. \ref{sus}, but without the presence
of any valley or shoulder.

\subsection{Entropy}
\label{entro}

The entropy, as a function of temperature, is shown in Fig. \ref{entro} for different values of
external field and exchange interaction $J_2$. In all cases, for high temperatures, the entropy
approaches the expected value $\mathcal{S}=\ln \omega$ with $\omega=32$, since in this
limit all states are equally probable. On the other hand, as $T$ goes to zero, one can see a
residual entropy occurring at the transition lines and regions of phase coexistence, compatible
with the results shown in Fig. \ref{pd}(b), with now $\omega$ being the degeneracy of the ground
state

\begin{figure}[h]
\includegraphics[scale=0.43]{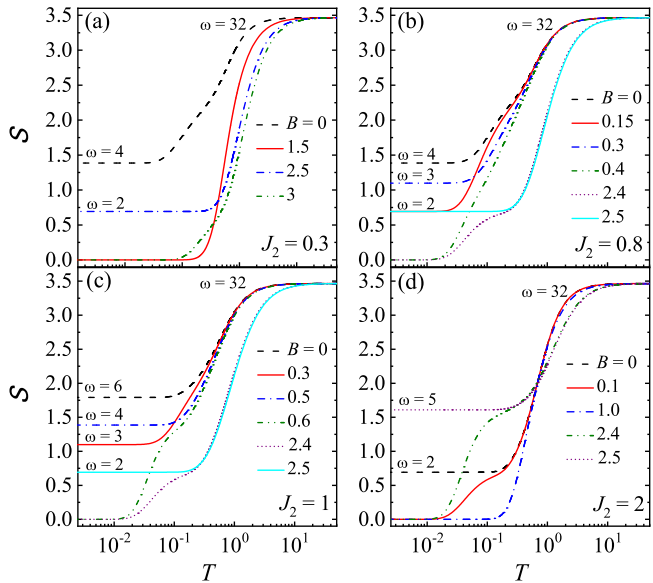}
\caption{\label{entro} (color online) Entropy $\mathcal{S}$ as a function of
 temperature $T$ (in logarithmic scale) for $J_3=-1$ and different values of the magnetic
 field $B$ and exchange interaction $J_2$. $\omega$ gives the degeneracy of the ground state
 (and also the number of states at high temperatures).}
\end{figure}

\begin{figure}[h]
\includegraphics[scale=0.42]{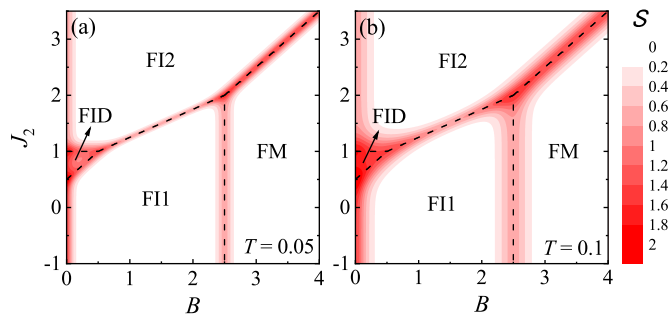}
\caption{\label{entropd} (color online) Density plot of the entropy $\mathcal{S}$, in a color gradient
 scale shown to the right, for two low temperatures, keeping in the background the phase diagram of
 Fig. \ref{pd}(b), using here only dashed lines for the transition boundaries.}
\end{figure}

In Fig. \ref{entropd} we have the entropy, in a gradient scale, for two different low temperatures
and the parameters of Fig. \ref{pd}(b), keeping in the background the zero temperature phase diagram.
We can see that the entropy at low temperatures is indeed higher near the lines of the phase
transitions in the region FID where several phases coexist.

Looking back to Fig. \ref{entro}, we can see that in (a)-(c) the entropy for $B=0$ is above the
entropy for greater values of $B$, implying that in this case one has a conventional
magnetocaloric effect. On the other hand, in (d) the entropy for $B=0$ is, in some intervals of $T$, below
the corresponding entropy of some greater external fields. As a consequence, in these regions we
have the inverse magnetocaloric effect.

\begin{figure}[h]
\includegraphics[scale=0.38]{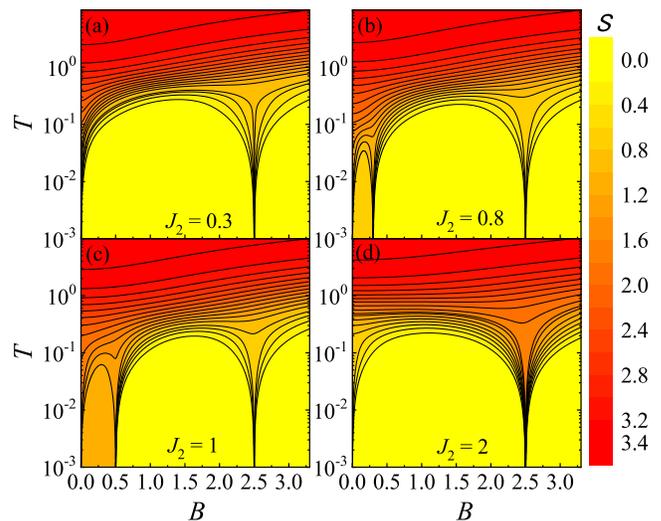}
\caption{\label{entroden} (color online) Density plot of the entropy, in a color gradient scale
 shown to the right, as a function the magnetic field $B$ and the temperature $T$ (in
 logarithmic scale) for $J_3=-1.0$  and different values of the exchange interaction $J_2$.
 Some lines of constant entropy are also plotted.}
\end{figure}

A more detailed view of the entropic behavior can be seen in Fig. \ref{entroden}. In this figure, we
have in fact the density plot of entropy, in a color gradient scale, as a function of $B$ and $T$
for $J_3=-1$. Some constant entropy lines are also highlighted in that figure and all lines meet,
for low temperatures, at the transition magnetic field values given in Fig. \ref{pd}(b).

\subsection{Magnetocaloric effect}
\label{mcef}

\begin{figure}[h]
\includegraphics[scale=0.4]{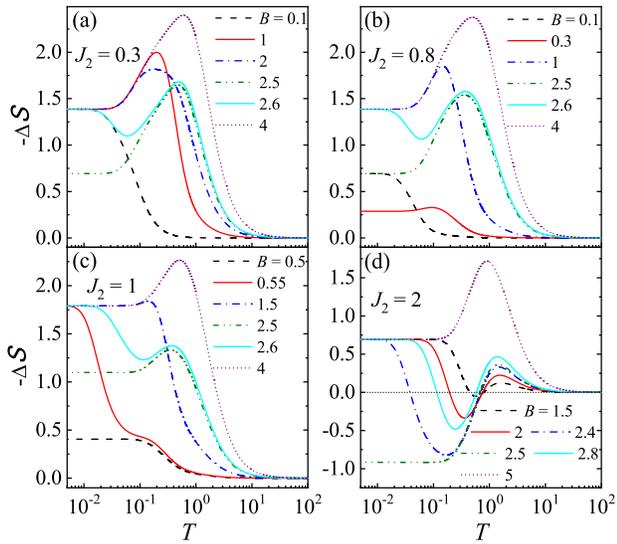}
 \caption{\label{entrochan} (color online) Isothermal change
 of the entropy $-\Delta \mathcal{S}$, as a function of the temperature $T$ (in logarithmic
 scale), for several values of the final magnetic field $B$ and considering a zero value for
 the initial magnetic field.}
\end{figure}

Although some aspects of the magnetocaloric effect could already be seen from the analysis of
Fig. \ref{entro}, more information can be obtained by computing the variation of the magnetic
entropy with the external field defined in Eq. (\ref{mce}). The change in the magnetic entropy
$-\Delta {\mathcal S}$ as a function of temperature for different values of the final magnetic field
is shown in Fig. \ref{entrochan}. In all cases we have considered zero initial magnetic field.
As discussed in the previous subsection, in Figs. \ref{entrochan}(a)-(c) only direct magnetic
caloric effect is present, because of the positiveness of $-\Delta \mathcal S$. However, in (d) we
clearly have an inverse magnetocaloric effect for some regions of temperature, since in this case
$-\Delta \mathcal S <0$. As for high temperatures the entropy is the same, regardless the value
of the external field, one has in this limit $-\Delta \mathcal S=0$ for any value of $B$. On the
other hand, in the limit of zero temperature $-\Delta \mathcal S$ is different from zero, due to
the ground state degeneracy at $B=0$.

\begin{figure}[h]
\includegraphics[scale=0.35]{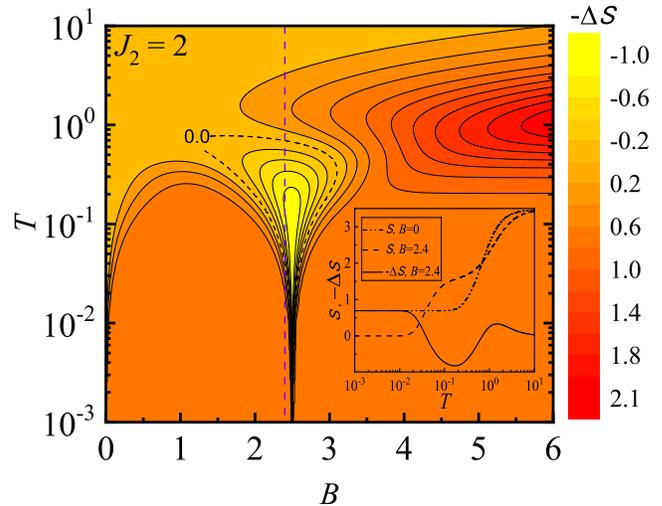}
\caption{\label{entrochanden} (color online) Density plot of the change in magnetic entropy $\Delta \mathcal S$,
 in a color gradient scale shown to the right, as a function the final magnetic field $B$ and the temperature
 $T$ (in logarithmic scale) for $J_3=-1.0$  and  $J_2=2$.
 Some lines of constant magnetic entropy change are also plotted. The dashed line corresponds to zero
 entropy variation. redThe inset shows the entropies and the magnetic entropy change for
 $B=2.4$ (close to the transition field $B=2.5$) as a function of temperature.}
\end{figure}

We can see from Fig. \ref{entrochan} that for small final fields the change in
the magnetic entropy presents a peak; for intermediate fields, generally about the phase transition
region, a minimum develops before the peak; and for still higher values of fields
only one peak is again observed. This is better seen in
Fig. \ref{entrochanden}, where we have the density plot of the change in the magnetic entropy
$-\Delta \mathcal S$ as a function of the final magnetic field $B$ and the temperature $T$
(in logarithmic scale) for $J_3=-1.0$  and  $J_2=2$. The inset in this figure shows the specific
case of the entropies and the magnetic entropy change for $B=2.4$ (close to the transition field
$B=2.5$) as a function of temperature. The minimum happens because the entropy for the
field in this region starts to increase before the zero field entropy and, in some cases, they
even cross, as shown in this inset. In the more detailed view of Fig. \ref{entrochanden}, the region
inside the dashed line corresponds to inverse magnetocaloric effect. For different values
of the exchange parameters than those of Fig. \ref{entro}, we only observe the
direct magnetocaloric effect.

\subsection{Specific heat}
\label{sph}

\begin{figure}[h]
\includegraphics[scale=0.4]{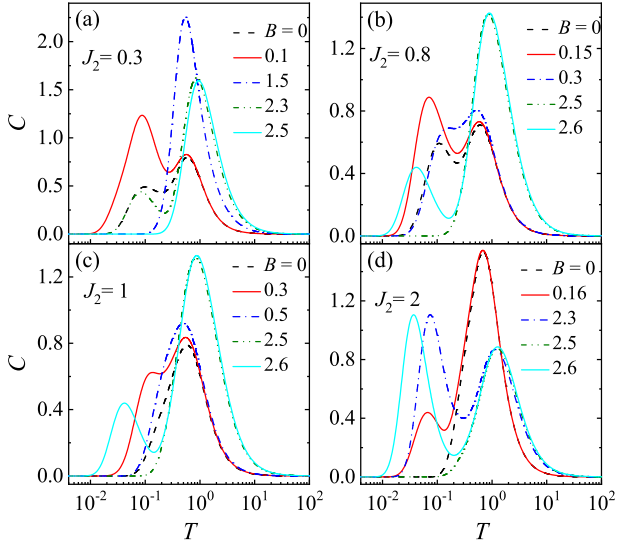}
\caption{\label{sh} (color online) Specific heat $C$
 as a function of temperature $T$ (in logarithmic scale) for various values of the magnetic
 field $B$ for $J_3=-1.0$.}
\end{figure}

Having computed the entropy as a function of temperature it is interesting to see the behavior
of the specific heat for these molecules. Fig. \ref{sh} shows $C$ as a function of temperature
for $J_3=-1$ and various values of external field and exchange interaction $J_2$. In all cases
the specific heat goes to zero in the limits of high and low temperatures, as should be the case
for any system having a finite energy spectrum. A Schottky behavior with only one maximum is noted for those values of external fields where the
entropy has no shoulder as function of temperature. The double peak topology, for certain values
of $B$, reflects exactly the shoulder present in the curves depicted in Fig. \ref{entro}.
The double peak topology for certain values of $B$ reflects the
shoulders present in the curves depicted in Fig. \ref{entro}. The experimental results of the
specific heat for the Cu-5NIPA compounds of Ref. \cite{nath} only show the low temperature
behavior of $C$. Experimental results for higher values of $T$ should evidence a kind of
Schottky peak for the fitted exchange interactions.

\section{Concluding remarks}
\label{conc}

A quantum spatially anisotropic spin-1/2 Heisenberg model, suitable for describing
Cu$_5$ pentameric molecules, in the presence of an external magnetic field applied
along the $z$-axis, has been studied through exact diagonalization of the Hamiltonian.
A wider range for the exchange interactions has been considered.
The behavior of the system at zero temperature has been determined through its energy
spectrum and a detailed analysis of the corresponding phase diagrams for different
coupling values. The model presents not only a rich phase diagram, but also
residual entropies at zero temperature. The thermodynamic properties at finite temperatures
have also been obtained by computing the magnetization,
susceptibility, entropy, magnetocaloric effect and specific heat. Depending on the values
of the exchange interactions these molecules can exhibit either direct or inverse magnetocaloric
effect. It is expected that the Hamiltonian (\ref{ham}), despite being a quite simple example
of a zero-dimensional quantum system, could also be applied to low dimensional quantum
models such as triangular chains.

\begin{acknowledgments}
The authors would like to thank CNPq (JT 159792/2019-3 and 163000/2020-4), Capes and FAPEMIG
for financial support.
\end{acknowledgments}

\appendix*
\section{}

The eigenvalues $\epsilon_i$ of the Hamiltonian \eqref{ham}, with $1\le i \le32$, can be
written as
\begingroup
\allowdisplaybreaks
\begin{align*}
  \varepsilon_1=&\frac{1}{2}J_1+\frac{1}{2}J_2-\frac{1}{2}J_3-\frac{5}{2}B\\
  \varepsilon_2=&\frac{1}{2}J_1+\frac{1}{2}J_2-\frac{1}{2}J_3+\frac{5}{2}B\\
  \varepsilon_3=&\frac{1}{2}J_1+\frac{1}{2}J_2-\frac{1}{2}J_3-\frac{3}{2}B\\
  \varepsilon_4=&\frac{1}{2}J_1+\frac{1}{2}J_2-\frac{1}{2}J_3+\frac{3}{2}B\\
  \varepsilon_5=&\frac{1}{4}J_1-\frac{1}{4}J_3-\frac{3}{2}B+\frac{1}{4}p\\
  \varepsilon_6=&\frac{1}{4}J_1-\frac{1}{4}J_3+\frac{3}{2}+\frac{1}{4}p\\
  \varepsilon_7=&\frac{1}{4}J_1-\frac{1}{4}J_3-\frac{3}{2}B-\frac{1}{4}p\\
  \varepsilon_8=&\frac{1}{4}J_1-\frac{1}{4}J_3+\frac{3}{2}B-\frac{1}{4}p\\
  \varepsilon_9=&-\frac{1}{4}J_1+\frac{1}{4}J_3-\frac{3}{2}B+\frac{1}{4}r\\
  \varepsilon_{10}=&-\frac{1}{4}J_1+\frac{1}{4}J_3+\frac{3}{2}B+\frac{1}{4}r\\
  \varepsilon_{11}=&-\frac{1}{4}J_1+\frac{1}{4}J_3-\frac{3}{2}B-\frac{1}{4}r\\
 \varepsilon_{12}=&-\frac{1}{4}J_1+\frac{1}{4}J_3+\frac{3}{2}B-\frac{1}{4}r\\
  \varepsilon_{13}=&\frac{1}{2}J_1+\frac{1}{2}J_2-\frac{1}{2}J_3-\frac{1}{2}B\\
  \varepsilon_{14}=&\frac{1}{2}J_1+\frac{1}{2}J_2-\frac{1}{2}J_3+\frac{1}{2}B\\
  \varepsilon_{15}=&\frac{1}{4}J_1-\frac{1}{4}J_3-\frac{1}{2}B+\frac{1}{4}p\\
  \varepsilon_{16}=&\frac{1}{4}J_1-\frac{1}{4}J_3+\frac{1}{2}B+\frac{1}{4}p\\
  \varepsilon_{17}=&\frac{1}{4}J_1-\frac{1}{4}J_3-\frac{1}{2}B-\frac{1}{4}p\\
  \varepsilon_{18}=&\frac{1}{4}J_1-\frac{1}{4}J_3+\frac{1}{2}B-\frac{1}{4}p\\
  \varepsilon_{19}=&-\frac{1}{2}J_1+\frac{1}{2}J_3-\frac{1}{2}B+\frac{1}{2}q\\
  \varepsilon_{20}=&-\frac{1}{2}J_1+\frac{1}{2}J_3+\frac{1}{2}B+\frac{1}{2}q\\
  \varepsilon_{21}=&-\frac{1}{2}J_1+\frac{1}{2}J_3-\frac{1}{2}B-\frac{1}{2}q\\
  \varepsilon_{22}=&-\frac{1}{2}J_1+\frac{1}{2}J_3+\frac{1}{2}B-\frac{1}{2}q\\
  \varepsilon_{23}=&-\frac{1}{4}J_1+\frac{1}{4}J_3-\frac{1}{2}B+\frac{1}{4}r\\
  \varepsilon_{24}=&-\frac{1}{4}J_1+\frac{1}{4}J_3+\frac{1}{2}B+\frac{1}{4}r\\
  \varepsilon_{25}=&-\frac{1}{4}J_1+\frac{1}{4}J_3-\frac{1}{2}B-\frac{1}{4}r\\
  \varepsilon_{26}=&-\frac{1}{4}J_1+\frac{1}{4}J_3+\frac{1}{2}B-\frac{1}{4}r\\
  \varepsilon_{27}=&x_1-\frac{1}{2}B\\
  \varepsilon_{28}=&x_1+\frac{1}{2}B\\
  \varepsilon_{29}=&x_2-\frac{1}{2}B\\
  \varepsilon_{30}=&x_2+\frac{1}{2}B\\
  \varepsilon_{31}=&x_3-\frac{1}{2}B\\
  \varepsilon_{32}=&x_3+\frac{1}{2}B\\
\end{align*}
\endgroup
where
\begin{align*}
  p=&\sqrt{(J_1+J_3)^2+4J_2^2},\\
  q=&\sqrt{(J_1+J_3)^2+J_2^2},\\
  r=&\sqrt{(J_1+J_3)^2+4J_2^2+8[J_1^2+J_3^2-J_2(J_1-J_3)]},
\end{align*}
and $x_i$, with $i=1,~2,~3$, are obtained as the solution of
\begin{align}
&8x^3+4(J_1+3J2-J3)x^2 \nonumber \\
&+2[J_2(2J_1-J_2+2J_3)-2(J_1+J_3)^2]x \nonumber \\
&-3J_2^2(J_1+J_2-J_3)=0 . \label{ce}
\end{align}

The corresponding eigenstates can also be obtained, but as they are rather lengthy they
are not reproduced here. Only the most relevant ones for the quantum phase diagrams are
given in the text.

\end{document}